\documentclass[english,a4paper,article,onecolumn,11pt]{memoir} 

\usepackage{babel} 
\usepackage[utf8]{inputenc} 
\usepackage[T1]{fontenc} 
\usepackage{lmodern} 
\usepackage{soul} 
\usepackage{microtype} 
\usepackage{graphicx} 
\usepackage{epstopdf} 
\usepackage{rotating} 
\usepackage{subfig} 
\usepackage[labelfont=bf]{caption} 
\usepackage[hidelinks,bookmarks,bookmarksopen,bookmarksdepth=2]{hyperref}

\captionstyle{\small} 
\captionnamefont{\small\bfseries} 
\usepackage{url} 

\setulmarginsandblock{3.5cm}{*}{1} 
\setlrmarginsandblock{1.6cm}{1.6cm}{*} 
\usepackage{amsmath,amsfonts,amssymb} 
\usepackage{mathtools} 
\usepackage{braket}

\newcommand{\abs}[1]{\left| #1 \right|} 

\title{\huge{How many atoms get excited when they decay?}}

\author{Philip Daniel Blocher and Klaus M\o lmer \\
Department of Physics and Astronomy, Aarhus University, DK 8000 Aarhus C, Denmark}
\date{(Dated: \today)}

\begin{document}
\maketitle
\begin{abstract}
We analyze the time evolution of a two-level system prepared in a superposition of its ground state and radiatively unstable excited state. We show that by choosing appropriate means of detection of the radiated field, we can steer the evolution of the emitter and herald its preparation in the fully excited state. We determine the probability for the occurrence of this "excitation during the decay" of a remote emitter.

\end{abstract}

\chapter{Introduction}
Preparation and control of the state of quantum systems is at the heart of a variety of applications in sensing, precision measurements, and quantum information science. While cooling and damping are instrumental mechanisms to prepare a known, pure state, e.g., an atomic ground state, measurements and their accompanying back action have turned out to be sometimes a more practical, efficient, and versatile tool. Numerous applications of heralding schemes as well as measurement and feedback schemes have been proposed and demonstrated in laboratories, and they are gradually becoming a significant component in our toolbox for state preparation and control. Thus, atomic ensembles \cite{Polzik} and remote pairs of individual ions \cite{Monroe} have been prepared in entangled states by detection of transmitted or spontaneously emitted photons, and a superconducting qubit has been prepared in its ground state by a state measurement (followed by a unitary inversion pulse, in case of an excited state outcome of the measurement) \cite{Dicarlo}. In conjunction with entangling interactions, heralding may substantially increase the degree of entanglement \cite{Anders1} or the fidelity of a quantum gate at the cost of an only near unity success probability  \cite{Anders2}.

In this article we consider the special case of a coherently excited two-level system that we may only access through the detection of radiation emitted during the spontaneous decay of the system into its ground state. We assume this decay to occur into a broadband continuum of radiation modes and, hence, we have no means to suppress the excited state decay rate $\Gamma$. The average excitation probability therefore follows the exponential decay law $P_e(t)= \pi_e \exp(-\Gamma t)$, where $\pi_e=P_e(0)$ is the initial excited state population. However, as the emitted quantum field is entangled with the state of the emitter, detection of the radiation causes a back action on this state. A similar effect exists in classical probability theory, where a light bulb may have an exponentially decaying survival probability, while any observed light bulb is in fact fully functional until it suddenly becomes defect. In both the classical and the quantum case, the observer cannot control or decide the outcome of the heralding measurement, but for quantum systems, the observer has the freedom to choose between different observables and hence to choose between different types of measurement back action on the emitter. This possibility was identified as a \textit{steering} property by Schr\"odinger \cite{SteeringSch}, and has been the subject of considerable recent theoretical and experimental interest \cite{SteeringWis,SteeringExp}.

Using a stochastic unraveling, we calculate the quantum state conditioned on the measurement of light emitted during the decay, and we assess the possibility that the emitter thus evolves from a partially to a fully excited state. We shall compare the situation of (i) quantum jumps associated with photon counting, (ii) quantum state diffusion associated with homodyne detection with a strong, classical local oscillator, and (iii) discrete jump-like behavior of the emitter associated with interference measurements with a weak local oscillator field.

The structure of this article is as follows. In Section 2, we recall the dynamics of quantum systems coupled to a broadband radiation reservoir and conditioned on detection of the signal emitted into the reservoir modes. In Section 3, we discuss how the measurement of the emitted signal combined with a local oscillator field of the correct amplitude can herald a quantum jump of the emitter into its excited state, and we characterize the probability for this process to occur. In Section 4, we extend the analysis to a strategy where the local oscillator amplitude is adapted according to previous measurement events, and in Section 5, we conclude with a summary and brief discussion of the relation between our work and the steering of quantum states.

\chapter{Quantum trajectories of a light emitting quantum system}
In this article we consider a two-state quantum system coupled by electric or magnetic dipole interaction to a freely propagating quantized radiation field. The excited state $\ket{e}$ has the energy $E_e$ above the ground state $\ket{g}$ with energy $E_g \equiv 0$, and we assume an emitter superposition state and no photons at time $t$, $\ket{\Psi(t)} = (a(t)\ket{g}+b(t)\ket{e})\otimes\ket{0}$. Following the Wigner-Weisskopf analysis of atomic decay, the wavefunction of the atom and the quantized radiation field at time $t + \text{d}t$ can then be written as
\begin{align}
\ket{\Psi(t+\text{d}t)} =& a(t)\ket{g}\otimes\ket{0} + b(t)\left(1 - \frac{\Gamma}{2}\text{d}t\right)\exp(-i E_e\text{d}t/\hbar)\ket{e}\otimes\ket{0}\nonumber\\
&+\sum_\lambda c_\lambda(t + \text{d}t)\exp(- i E_\lambda\text{d}t/\hbar)\ket{g}\otimes\ket{1_\lambda}\label{eq:SystemReservoirCouplingPerturbation}
\end{align}
where the first two terms represent the state of the emitter with no photons present in the quantized field, while the last term represents the ground state emitter in the presence of a one-photon state, which has been expanded on, e.g., a set of traveling wave modes $\ket{1_\lambda}$ with photon energies $E_\lambda$.

According to (\ref{eq:SystemReservoirCouplingPerturbation}), the atomic coherences and populations decay due to the coupling to the quantized field. If we assume that previously emitted photons do not affect the future atomic dynamics (Markov approximation), we can derive a master equation for the system density matrix $\boldsymbol{\rho}$, according to which the excited state population decays exponentially, $\dot{\rho}_{ee}=-\Gamma\rho_{ee}$, while, in a frame rotating at $(E_e-E_g)/\hbar$, the coherences decay according to $\dot{\rho}_{eg}=-\frac{\Gamma}{2}\rho_{eg},\ \rho_{ge}=\rho_{eg}^*$. The last term in (\ref{eq:SystemReservoirCouplingPerturbation}) represents the return of the population decay of the excited state into the ground state, $\dot{\rho}_{gg}=+\Gamma\rho_{ee}$.

\section{Quantum jumps}
While the rate equations for the density matrix elements describe the average, unobserved dynamics of the emitter, detection of the emitted radiation heralds a different, stochastic evolution \cite{Carmichael,MCWF}. This is most easily appreciated by considering the effect of photon counting on the state described by (\ref{eq:SystemReservoirCouplingPerturbation}) at time $t+dt$: Detection of no photon causes a projection onto the first two components of (\ref{eq:SystemReservoirCouplingPerturbation}) (with no photons in the field) and effectively leads to a continuous reduction of the excited state amplitude. Detection of a photon similarly projects the system onto the last term of (\ref{eq:SystemReservoirCouplingPerturbation}), i.e. the emitter is subject to a quantum jump into the ground state, accompanied by the annihilation of the photon in case of normal photon counting.

\begin{figure}[h]
  	\begin{minipage}[c]{0.45\textwidth}
  		\centering
  		\includegraphics[scale=1]{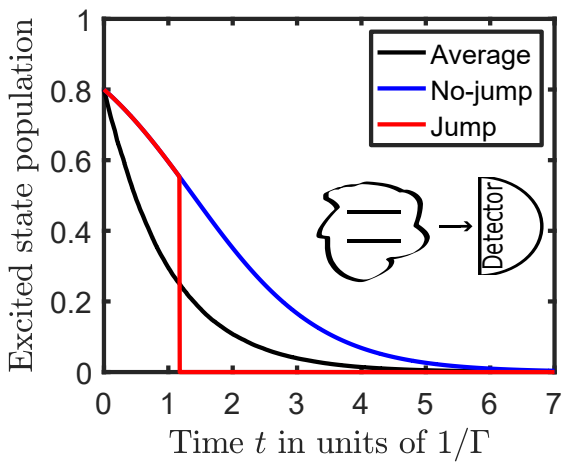}
  	\end{minipage}\hfill
  	\begin{minipage}[c]{0.5\textwidth}
		  	\caption{\textbf{Main figure:} The excited state population of a two-level system subject to the no-jump dynamics (upper, continuous blue curve), subject to a quantum jump at $t=1.2 \Gamma^{-1}$ (stepwise continuous red curve), and the average population over 20,000 simulations (lower, black curve), in perfect agreement with the exponential decay $\pi_e\exp(-\Gamma t)$. \textbf{Inset:} Schematic diagram of the photon counting monitoring of the decay of the emitter.}
		  	\label{fig:QuantumJumps}
  	\end{minipage}
\end{figure}

By considering the full time evolution of an emitter prepared in the initial state $a(0)\ket{g}+b(0)\ket{e}$, we arrive at the following simple dynamics: during time intervals with no photon detection, the excited state amplitude decays exponentially as $b\exp(-\Gamma t/2)$ and, after renormalization of the state, the conditional excited state population takes the form $P^\text{cond}_{e}(t)=\pi_e\exp(-\Gamma t)/(\pi_g+ \pi_e\exp(-\Gamma t))$, where $\pi_g = 1-\pi_e$ is the initial ground state population. The detection of a photon in a time interval $dt$ has the time dependent probability $dP(t) = P^\text{cond}_{e}(t)\Gamma dt$, and one can show that a fraction $\pi_e$ ($\pi_g$) of the emitters reach the ground state with (without) a quantum jump. The no-jump and the different jump trajectories precisely average to yield the conventional exponential decay curve, but in none of the trajectories will the excited state population exceed its initial value. Figure~\ref{fig:QuantumJumps} demonstrates both a single trajectory containing a jump, a single no-jump trajectory, as well as the average of 20,000 trajectories for such a system.

\section{Homodyne detection}
Photon counters yield discrete count events with a rate proportional to the intensity of the field from the emitter. By beating the emitter signal with a strong, coherent local oscillator field, the intensity of the interfering fields acquires components proportional to the emitted field amplitude. To describe this situation theoretically, we apply the schematic shown in the inset of Fig.~\ref{fig:moderateAlphaHomodyne}, where the emitter signal and a coherent state with amplitude $\alpha$ are combined on a beam splitter and the two output fields are subject to photon counting.

The coherent state of radiation $\ket{\alpha}$ is an eigenstate of the annihilation operator $\hat{a}$, with $\hat{a}\ket{\alpha} = \alpha \ket{\alpha}$, and the photon flux in the beam is given by the mean value $\braket{\hat{a}^\dagger \hat{a}} = \abs{\alpha}^2$. The annihilation of a photon from the emitter signal is associated with a quantum jump of the emitter, as described by the operator $\hat{C}=\sqrt{\Gamma}\ket{g}\bra{e}$, where the pre-factor leads to the predicted counting rate $\langle\hat{C}^\dagger \hat{C}\rangle = \Gamma\rho_{ee}$.

As we are mixing the fields on a beam splitter of intensity transmission coefficient $\mu$, the annihilation of a photon in either of the output ports is governed by a superposition of the annihilation operators of the input modes,
\begin{align}
\hat{C}_\text{A} =& \sqrt{(1-\mu)}\hat{a} + \sqrt{\mu}\sqrt{\Gamma}\ket{g}\bra{e},\nonumber\\
\hat{C}_\text{B} =& \sqrt{\mu}\hat{a} - \sqrt{(1-\mu)}\sqrt{\Gamma}\ket{g}\bra{e}, \label{eq:HomodyneOperators}
\end{align}
where we have chosen real reflection and transmission amplitudes without loss of generality.

Acting on an arbitrary product state $(a(t)\ket{g}+b(t)\ket{e})\otimes\ket{\alpha}$ of the emitter and local oscillator, these operators yield the probabilities $dP_{A(B)} = \braket{\hat{C}_{A(B)}^\dagger \hat{C}_{A(B)}} dt$ for photon detection in the two detectors.  The (un-normalized) state resulting as the outcome of a click in detector A reads
\begin{equation}
\ket{\Psi}_{A} = \hat{C}_{A}\ket{\Psi} =  \left[\sqrt{(1-\mu)}\alpha a(t) + \sqrt{\mu} \sqrt{\Gamma} b(t) \right]\ket{g} + \sqrt{(1-\mu)}\alpha b(t) \ket{e}, \label{eq:DetectorAJump}
\end{equation}
and we may find the resulting state for a click in detector B in the same manner. Contrary to the quantum jump into the ground state by direct photon counting, performing a measurement on the interference signal from the emitter and coherent local oscillator allows the emitter quantum state to retain a non-vanishing excited state population.

\subsection{Homodyne detection with a strong and a weak local oscillator}
\begin{figure}[h]
	\centering
	\includegraphics[scale=1]{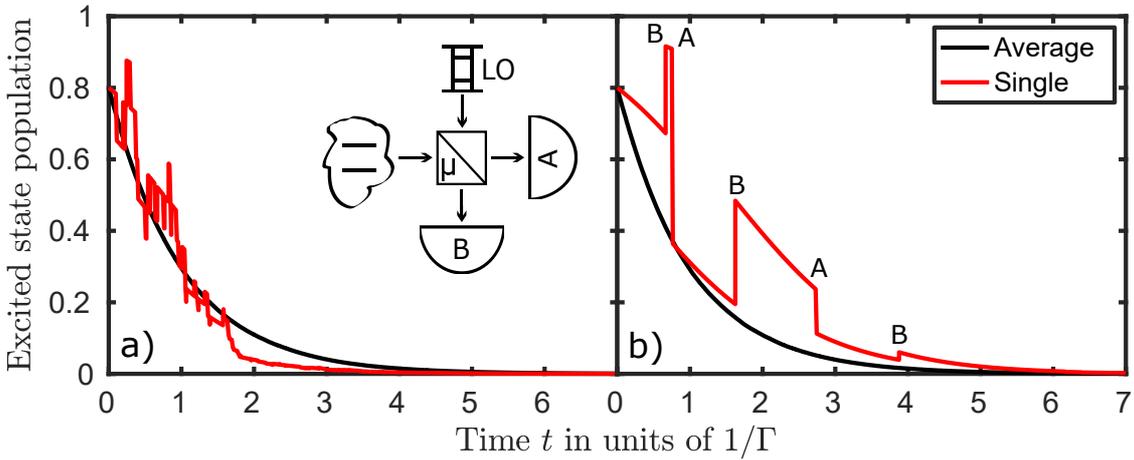}
	\caption{\textbf{a):} The excited state population for a two-level system subject to counting of photons after interference on a lossless beam splitter with a coherent state with amplitude $\alpha=5\sqrt{\Gamma}$. The red curve shows how detection events lead to a small decrease or increase of the excited state population which in the limit of large $\alpha$ becomes a continuous stochastic process. \textbf{b):} The excited state population for a two-level system subject to counting of photons after interference on a lossless beam splitter with a coherent state with amplitude $\alpha=\sqrt{\Gamma}$. The red curve shows how detection events lead to a finite, discontinuous decrease or increase of the excited state population. For both a) and b), the lower, black curve is the average over 20,000 trajectories. \textbf{Inset:} Schematic diagram of the emitter plus coherent LO setup.}
	\label{fig:HomodyneLargeAlpha}
	\label{fig:moderateAlphaHomodyne}
\end{figure}

For a strong local oscillator, the evolution is dominated by the terms proportional to $\alpha$ in (\ref{eq:DetectorAJump}) and in the similar equation for detector $B$. These terms preserve the amplitudes of the two-level system, while the term independent of $\alpha$ differs in sign (and magnitude if $\mu \neq 0.5$) depending on which detector clicks. The resulting dynamics becomes a diffusion process in Hilbert space and can also be represented by a stochastic differential equation of Ito or Stratonovich form \cite{WisMil}. We shall not write this equation here but merely note that if the initial state amplitudes and the local oscillator field amplitude are all real, this will remain true for the entire evolution of the system. The two-level emitter dynamics will therefore follow a great circle on the Bloch sphere, where the decay on average drives the system towards the south pole (ground state) while the stochastic measurement back action causes diffusion of the lattitude coordinate, see Fig.~\ref{fig:HomodyneLargeAlpha}.a. This diffusion may both randomly excite and de-excite the emitter, and, for an initial superposition state with real amplitudes and $b(0)\neq 0$, it was shown in \cite{Bolund} that there is a non-vanishing probability to reach the fully excited state $b(t)=1$ at some later time during the homodyne detection of the emitted radiation. This probability, $P_\text{Hom}$, follows from Eqs. (13, 14) in \cite{Bolund} and is shown as a function of $\pi_e$  in Fig.~\ref{fig:DynamicProtocol}.

For a weak local oscillator the observed system dynamics are of a discrete jump character, where both upward and downward jumps occur according to which detector clicks and the state amplitudes of the emitter. These dynamics are sketched in Fig.~\ref{fig:HomodyneLargeAlpha}.b, where the respective jumps have been marked with A (B) to indicate a photon detection in detector A (B).

\chapter{Homodyne detection with a weak, adaptive local oscillator}
\begin{figure}[h]
	\begin{minipage}[c]{0.4\textwidth}
		\centering
		\includegraphics[scale=1]{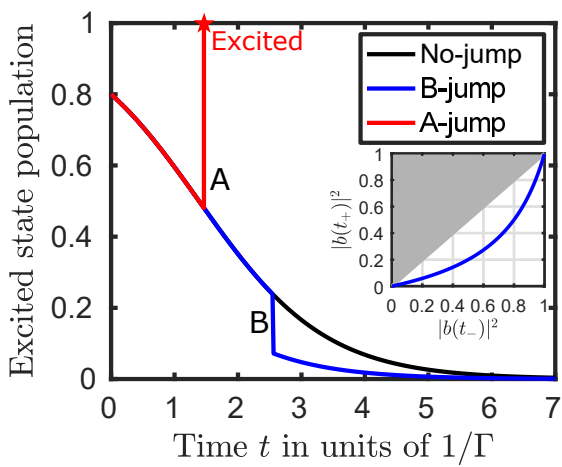}
	\end{minipage}\hfill
	\begin{minipage}[c]{0.55\textwidth}
	    \caption{\textbf{Main figure:} The excited state population for a two-level system monitored by the adaptive detection scheme of section 3. The figure displays the excited state population of the system subject to a detector A quantum jump at $t = 1.4\Gamma^{-1}$ (red stepwise continuous curve), subject to a single detector B quantum jump at $t = 2.5\Gamma^{-1}$ (blue stepwise continuous curve), as well as subject to no-jump dynamics (black continuous curve). \textbf{Inset:} The excited state population $|b(t_+)|^2$ resulting from a click in detector B at time $t$ as a function of the pre-jump excited state population $|b(t_-)|^2$.}
		\label{fig:AdaptiveDetectionTrajectory}
	\end{minipage}
\end{figure}

We have seen that the state of the system after detection of a photon in one of the detectors depends on the choice of local oscillator amplitude $\alpha$: If $\abs{\alpha}$ is very large, the state changes frequently in infinitesimal diffusive steps, Fig.~\ref{fig:HomodyneLargeAlpha}.a, while if $\abs{\alpha}$ is vanishing, the backaction amounts to a jump into the ground state, as in Fig.~\ref{fig:QuantumJumps}. We note that we may also choose the strength $\alpha$, such that the detection of a photon in detector A leads to a fully excited state of the emitter. This merely requires that the ground state amplitude in (\ref{eq:DetectorAJump}) vanishes, i.e., we have to adapt the value of $\alpha$ to the state of the emitter:
\begin{equation}
\alpha(t) = -\sqrt{\Gamma}\sqrt{\frac{\mu}{1-\mu}} \frac{b(t)}{a(t)},
\label{eq:DynamicAlpha}
\end{equation}
where we recall that the state amplitudes $a(t)$ and $b(t)$ are time dependent quantities. The upper red curve in Fig. \ref{fig:AdaptiveDetectionTrajectory} shows the excited state population of an emitter, conditioned on no counting events until a detector A count at $t=1.4 \Gamma^{-1}$ heralds the preparation of a fully excited emitter state. 

While a detection event in detector A leads to the excited state, we are not guaranteed to obtain this outcome. In fact, detector B may also register a photon, in which case the (un-normalized) conditioned state, assuming (\ref{eq:DynamicAlpha}), becomes
\begin{equation}
\hat{C}_\text{B}\ket{\Psi(t)} = \left[\frac{-\sqrt{\Gamma}}{\sqrt{1-\mu}}b(t)\right]\ket{g} + \left[\frac{-\mu\sqrt{\Gamma}}{\sqrt{1-\mu}}\frac{b^2(t)}{a(t)}\right]\ket{e}.
\label{eq:BJumpState}
\end{equation}
This state is less excited than it was before the detection. The lower, blue curve in Fig. \ref{fig:AdaptiveDetectionTrajectory} shows an example of the excited state population conditioned on no counting events, until a detector B count at $t=2.5 \Gamma^{-1}$ causes a quantum jump which yields a lower excited state population. The inset of Fig. \ref{fig:AdaptiveDetectionTrajectory} demonstrates that the excited state population $|b(t_+)|^2$ after a B detection at time $t$ is always smaller than the excited state population $|b(t_-)|^2$ before the detection event.

\section{Probability of obtaining a fully excited state by an adaptive homodyne detection scheme}
\subsection{Limits to success}
With how high a probability will the adaptive homodyne detection scheme herald the preparation of the two-level emitter in the excited state? As our measurement is based on the detection of radiation dissipated by the emitter, the success probability of our adaptive scheme must be smaller than the initial excited state population $\pi_e$, which is, in turn, the probability to obtain the excited state in an initial projective measurement on the emitter itself.

To analyze our protocol in more detail, we shall now determine the probability that no detection events take place during monitoring of the system and hence that it evolves continuously into the ground state.
This probability is evaluated by propagating the system as if subject to a non-Hermitian Hamiltonian, $H_\text{eff}= -\frac{i\hbar}{2}( \hat{C}^\dagger_\text{A}\hat{C}_\text{A}+\hat{C}^\dagger_\text{B}\hat{C}_\text{B})$ \cite{MCWF}, giving rise to the equation of motion for the \textit{un-normalized} density matrix $\boldsymbol{\rho}$,
\begin{equation}
\dot{\boldsymbol{\rho}}(t) = -\frac{1}{2}\left\{\hat{C}_\text{A}^\dagger\hat{C}_\text{A} + \hat{C}_\text{B}^\dagger\hat{C}_\text{B},\boldsymbol{\rho}(t)\right\}
= -
\begin{pmatrix}
\abs{\alpha(t)}^2 \rho_{gg}(t) && \left(\abs{\alpha(t)}^2+\frac{\Gamma}{2}\right)\rho_{ge}(t) \\
\left(\abs{\alpha(t)}^2+\frac{\Gamma}{2}\right)\rho_{eg}(t) && \left(\abs{\alpha(t)}^2 + \Gamma\right) \rho_{ee}(t)
\end{pmatrix},
\label{eq:DynamicLindblad}
\end{equation}
with the initial density matrix elements $\rho_{gg}(t=0)=\pi_g$, $\rho_{ee}(t=0)=\pi_e$, $\rho_{ge}(t=0)=\rho_{eg}(t=0)=\sqrt{\pi_e\pi_g}$.

By inserting $|\alpha(t)|^2=\frac{\mu\Gamma}{1-\mu} \frac{\rho_{ee}(t)}{\rho_{gg}(t)}$ from (\ref{eq:DynamicAlpha}), we obtain from the second derivative of $\rho_{gg}(t)$ the closed differential equation
\begin{equation}
\ddot{\rho}_{gg}(t) = -\frac{\mu\Gamma}{1-\mu}\dot{\rho}_{ee}(t)
= \frac{\dot{\rho}_{gg}{}^2(t)}{\rho_{gg}(t)} - \Gamma \dot{\rho}_{gg}(t). \label{eq:DifferentialEquationDynamic}
\end{equation}
This equation has the solution
\begin{equation}
\rho_{gg}(t) = \pi_g \exp\left(\frac{-\mu}{1-\mu}\frac{\pi_e}{\pi_g} \Big(1-\exp\left(-\Gamma t\right)\Big)\right),
\label{eq:EoMGround}
\end{equation}
and leads to
\begin{equation}
\rho_{ee}(t) = \pi_e \exp\left(\frac{-\mu}{1-\mu}\frac{\pi_e}{\pi_g} \Big(1-\exp\left(-\Gamma t\right)\Big)\right)\exp(-\Gamma t),
\label{eq:EoMExcited}
\end{equation}
for the un-normalized $\rho_{gg}(t)$ and $\rho_{ee}(t)$.

The probability $P_0(t)$ that no jump has been detected until time $t$ is given by the trace of the time-dependent, un-normalized, no-jump density matrix:
\begin{equation}
P_0(t) = \text{Tr}(\boldsymbol{\rho}(t)) = \Big(\pi_g + \pi_e\exp(-\Gamma t)\Big)\exp\left(\frac{-\mu}{1-\mu}\frac{\pi_e}{\pi_g} \Big(1-\exp\left(-\Gamma t\right)\Big)\right),
\end{equation}
and therefore we find the probability of no detection at all during the decay of the emitter to be
\begin{equation}
P_0 = \lim_{t\rightarrow\infty} P_0(t) = \pi_g \exp\left(\frac{-\mu}{1-\mu}\frac{\pi_e}{\pi_g}\right). \label{eq:nojumpprob}
\end{equation}
This probability is non-vanishing because the adapted local oscillator amplitude (\ref{eq:DynamicAlpha}) is gradually reduced as the excited state amplitude becomes smaller.

The probability that we will detect a photon in detector A is bound to be less than or equal to $1-P_0$, and we plot this upper limit to the success probability of our protocol as the upper blue curve in Fig.~\ref{fig:DynamicProtocol}. We observe in the figure that this constraint is weaker than the energy conservation constraint shown by the shaded area above $P=\pi_e$ in the figure.

\subsection{A single A-jump}
The probability $dP_A(t)$ that we will detect no photon in any of the detectors until time $t$ followed by the desired A detection in the next infinitesimal time step $dt$ is given by
\begin{equation}
dP_A(t) = \text{Tr}\left(\boldsymbol{\rho}(t)\right)\times
\frac{\text{Tr}\left(\boldsymbol{\rho}(t)\hat{C}_\text{A}^\dagger\hat{C}_\text{A}\right)}
{\text{Tr}\left(\boldsymbol{\rho}(t)\right)}\text{d}t = \mu\Gamma\frac{\rho_{ee}{}^2(t)}{\rho_{gg}(t)}\,\text{d}t
\label{eq:DynamicProbabilityAInfinity}
\end{equation}
with the expressions for $\rho_{ee}(t)$ and $\rho_{gg}(t)$ given in Eqs.~(8, 9).

\begin{figure}[h]
	\begin{minipage}[c]{0.45\textwidth}
		\centering
		\includegraphics[scale=1]{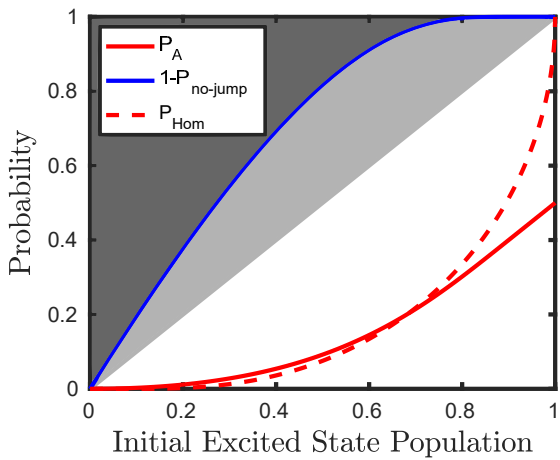}
	\end{minipage}\hfill
	\begin{minipage}[c]{0.5\textwidth}
		\caption{The lower solid red curve in the figure shows the probability $P_A$ that the emitter jumps into the fully excited state due to an A detector event at any time following a no-time evolution since the coherent excitation of the emitter with initial excited state population $\pi_e$. We assume a 50:50 beam splitter with $\mu=0.5$. The success probability is excluded from the shaded areas defined by the energy conservation $P = \pi_e$ and by the no-jump probability (\ref{eq:nojumpprob}). The lower dashed curve shows the probability to reach the fully excited state conditioned on balanced homodyne detection with a strong oscillator field \cite{Bolund}.}
		\label{fig:DynamicProtocol}
	\end{minipage}
\end{figure}

The total probability of an A-detection happening without any preceding detection events is the integral of this expression from time zero to infinity,
\begin{equation}
P_A = \int_0^\infty \mu\Gamma \frac{\rho_{ee}{}^2(t)}{\rho_{gg}(t)}\,\text{d}t
\label{eq:P1}
\end{equation}
This integration is evaluated analytically in the appendix (Eq.~(\ref{Appeq:P_A})) and the result is shown as the red solid curve in Fig.~\ref{fig:DynamicProtocol}. 

\begin{figure}[h]
	\begin{minipage}[c]{0.6\textwidth}
		\centering
		\includegraphics[scale=1]{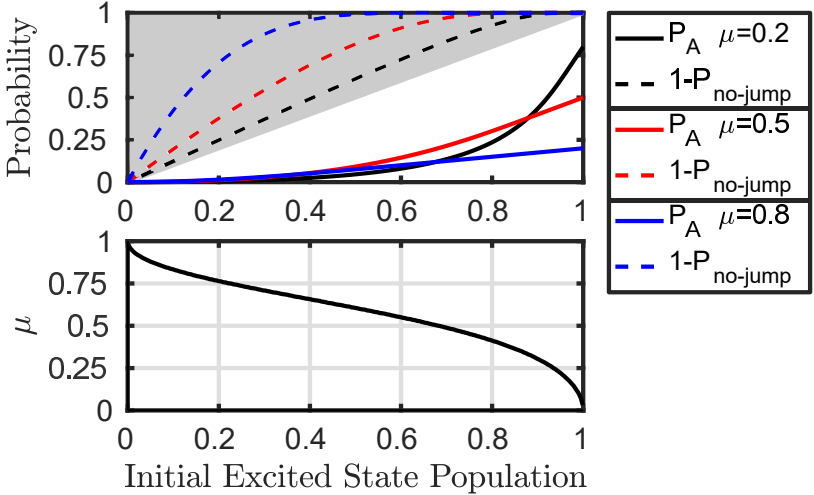}
	\end{minipage}\hfill
	\begin{minipage}[c]{0.35\textwidth}
		\caption{\textbf{Upper panel:} The solid curves show the probability that the emitter jumps into the fully excited state due to an A detector event at any time following a no-jump evolution since the initial preparation with excited state populations $\pi_e$. The shaded area and the dashed lines show the probabilities excluded by energy conservation and by the no jump probability, respectively. From below (above) at $\pi_e=1.0(0.2)$, the solid (dashed) curves correspond to $\mu=0.8,\ 0.5$ and $0.2$, respectively. \textbf{Lower panel:} The black curve yields the choice of $\mu$ that maximizes the probability $P_A$ of an A-jump occuring for a given initial excited state population $\pi_e$.}
  		\label{fig:DynamicProtocolVariableMu}
	\end{minipage}
\end{figure}

In Fig.~\ref{fig:DynamicProtocol}, the beam splitter was chosen to be balanced, i.e. $\mu = 0.5$. However, along with the local oscillator field amplitude $\alpha$, $\mu$ may also be adapted to yield the highest possible success probability of our scheme. The beam splitter transmissivity appears in the expressions for the no-jump and the A-jump probabilities, and the upper panel of Fig.~\ref{fig:DynamicProtocolVariableMu} shows the A-jump probability and the excluded probabilities as functions of $\pi_e$ for different values of $\mu$. As the figure shows, for a larger (smaller) initial excitation, a small (large) transmissivity is optimal. The lower panel shows the value of $\mu$ that yields the highest A-jump probability for a given initial excited state population; this value of $\mu$ is found from the equation for $P_A$ presented in the appendix. The figure is suggestive of a protocol that adapts $\mu$ with time. This may be possible by using e.g. a Mach-Zender interferometer with adjustable path lengths as a beam splitter. While such optimization could be an interesting topic for further study, we shall for simplicity consider only fixed values of $\mu$ in the remaining sections of this article.

\chapter{Adaption to multiple detection events}
In Figs.~\ref{fig:DynamicProtocol} and~\ref{fig:DynamicProtocolVariableMu} we observe a rather large gap between the success probability (\ref{eq:P1}) and the upper limit inferred by the no-jump probability (\ref{eq:nojumpprob}) and by energy conservation. This gap is associated with the probability for other possible detector outcomes, namely the detection of one or more clicks by detector B.

In the same manner as we calculated the probability that the first detection is by detector A at time t, we may also obtain the probability that the first detection is by detector B. Rather than discarding the ensuing dynamics, we can adapt the local oscillator amplitude such that a subsequent A detection will yield the excited state. 

\begin{figure}[h]
	\centering
	\includegraphics[scale=1]{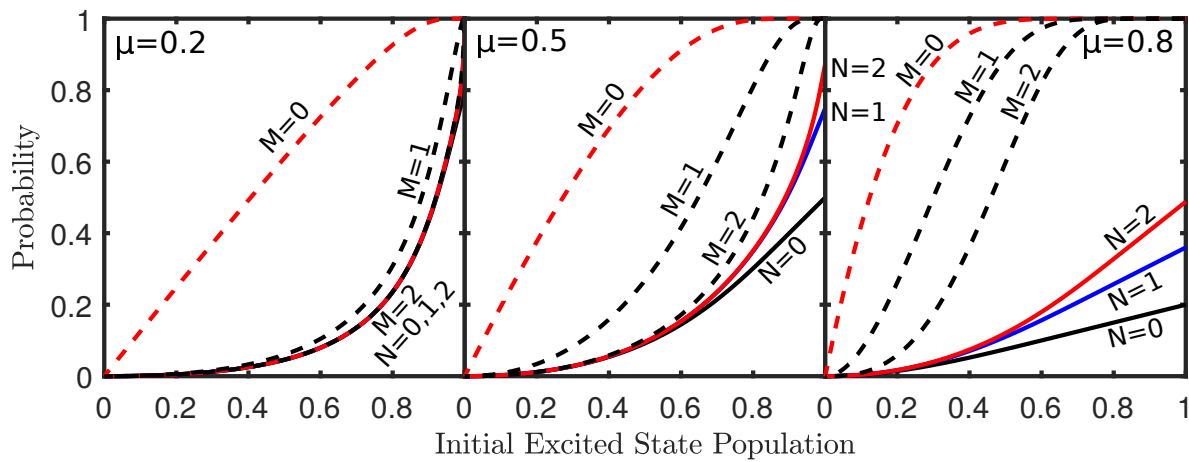}
	\caption{Probabilities for different event sequences, shown as functions of the initial  excited state population $\pi_e$ for $\mu=0.2,\ 0.5$ and $0.8$ (left to right panels). Solid lines show $P_N=\sum_{n=0}^N P_{\text{B}^n \text{A}}$ for event sequences ending with an A-jump and dashed lines show $Q_M=1-\sum_{m=0}^M P_{\text{B}^m 0}$ for sequences with no A-jump, with $N$ and $M$ indicated in the figure, see text.}
	\label{fig:DynamicMu}
\end{figure}

Following the analysis in the previous subsections, we are able to calculate the probability that no jumps occur after the B-detection and, by integrating over the time of the B-detection, the total probability $P_{\text{B}0}$ of reaching the emitter ground state after a single B-jump. We may proceed along the same line of thought and evaluate the probability $P_{\text{B}^m0}$ of reaching the ground state after $m$ B-jumps and no further jumps. When these probabilities are subtracted from unity in a cumulative manner, $Q_M \equiv 1-\sum_{m=0}^M P_{\text{B}^m0}$, they yield a increasingly tighter constraint on the remaining probability to herald the perfectly excited emitter state, see the dashed curves in Fig.~\ref{fig:DynamicMu}. We observe in the figure that the inclusion of a single B-jump ($Q_1$) strengthens the constraint set by the initial excitation $\pi_e$ and energy conservation.

Similar to the previous subsection, we may also evaluate the probability $P_{\text{B}^n\text{A}}$ that, after $n$ (none or several) B-jumps, the system undergoes an A-jump and heralds the fully excited state. These probabilities all contribute to the success probability of the protocol. We have evaluated expressions for these probabilities in terms of integrals over the heralding event times up to $n=2$ as well as for $m=2$ B-jumps followed by no-jump dynamics. The rather lengthy expression for $P_\text{BA}$ is given as (\ref{Appeq:P_BA}) in the appendix and presented in Fig.~\ref{fig:DynamicMu}. We observe that the gap between the achievable and the excluded probabilities gradually closes when we include more terms in the accumulated probability $P_N \equiv \sum_{n=0}^N P_{\text{B}^n\text{A}}$ and in the excluded probability $Q_M$, in particular for the lower beam splitter transmissivity.

\chapter{Discussion}
In summary, we have studied how the monitoring of the excitation of a broadband radiation reservoir affects the possible dynamical evolution trajectories of a quantum emitter. The monitoring does not change the average decay of the system, but it determines the type of evolution of the system, as illustrated by the possibility to sometimes herald a fully excited state. The use of measurements to herald specific states of a quantum system sees widespread application, and we have in this article demonstrated how one may optimize the success probability of remotely preparing an excited state by adapting the measurement scheme during continuous probing. We illustrated this idea by homodyne detection using a weak local oscillator with an adaptive coherent field strength, and, as we showed, the beam-splitter parameter governing the mixing of the emitted field and the local oscillator field may be varied to further improve the scheme.

We have assumed perfect collimation, lossless transmission and unit detection efficiency of the emitted signal. For imperfect homodyne (and heterodyne) detection, the quantum trajectories of a decaying system explore mixed states \cite{Exp1,Exp2}, and one may not herald the pure excited state with certainty. In that situation, one may instead investigate the probability that the heralded mixed state has an excited state population exceeding a certain critical value, e.g., above the initial excitation of the system.

Finally, let us recall that our protocol relies on the entanglement of the emitter and the spontaneously emitted radiation, and that the apparent paradoxical excitation of the emitter by the monitoring of its decay reflects the non-local measurement back action, as criticized by Einstein and Schr\"odinger in the early days of quantum mechanics. Schr\"odinger associated the term \textit{steering} \cite{SteeringSch} to the way that an observer affects the dynamics of one system by the choice of measurement the observer performs on another, possibly remote one. As demonstrated by Wiseman et al \cite{SteeringWis}, the steering property can be quantified and unambiguously verified in experiments. As also shown by Wiseman et al \cite{JumpsWis,SteeringDet}, one may quantify different measurement schemes, e.g. counting and homodyne detection, by their ability to steer the emitter state, and how that ability is degraded for reduced detector efficiencies. The results in \cite{SteeringDet} assume ergodic dynamics, and we may ask under which conditions the transient evolution, where only a single quantum of excitation is available for detection, allows for the verification of steering of the emitter dynamics.

\newpage\thispagestyle{empty}
$ $
\appendix
\chapter{Appendix}
\label{appendix:Probabilities}
\renewcommand{\theequation}{A.\arabic{equation}}
\setcounter{equation}{0}
Here we will carry out the integral in (\ref{eq:P1}) for the probability to obtain a single A detection event. By inserting the expressions of (\ref{eq:EoMGround}) and (\ref{eq:EoMExcited}), we find that
\begin{align}
P_\text{A}
=& \int_0^\infty\text{d}t\, \mu\Gamma \frac{\rho_{ee}{}^2(t)}{\rho_{gg}(t)}
= \mu \Gamma \int_0^\infty\text{d}t\, \frac{\pi_e{}^2\,\exp\left(\frac{-2\mu}{1-\mu}\frac{\pi_e}{\pi_g}\right) \exp(-2\Gamma t) \exp\left(\frac{2\mu}{1-\mu}\frac{\pi_e}{\pi_g} \exp(-\Gamma t)\right)}{\pi_g \exp\left(\frac{-\mu}{1-\mu}\frac{\pi_e}{\pi_g}\right) \exp\left(\frac{\mu}{1-\mu}\frac{\pi_e}{\pi_g}\right)} \nonumber\\
=& \mu \Gamma \frac{\pi_e{}^2}{\pi_g} \exp\left(\frac{-\mu}{1-\mu}\frac{\pi_e}{\pi_g}\right) \int_0^\infty\text{d}t\, \exp(-2\Gamma t)  \exp\left(\frac{\mu}{1-\mu}\frac{\pi_e}{\pi_g}\exp(-\Gamma t)\right)\nonumber \\
=& \frac{\mu \pi_e{}^2}{\pi_g}\, \left(\frac{\exp\big(A\big)}{A^2} + \frac{1-\exp\left(A\right)}{A^3}\right),
\label{Appeq:P_A}
\end{align}
where we have defined 
\begin{equation}
 A = \exp\left(\frac{\mu}{1-\mu}\frac{\pi_e}{\pi_g}\right).
\end{equation}

In the same manner as above we may determine the probability of a B detection event followed by an A detection event, $P_\text{BA}$. This involves an integral over the corresponding two times $t_B$ and $t$ for respectively the B-jump and the A-jump. Letting $\rho_{B,ee}(t)$ and $\rho_{B,gg}(t)$ denote density matrix elements evolved to the time $t$ conditioned on the outcome of the B-jump at time $t_B$ as given by (\ref{eq:BJumpState}), we find that
\begin{align}
P_\text{BA} =& \int_0^\infty\text{d}t\,  \mu\Gamma \frac{\rho_{B,ee}{}^2(t)}{\rho_{B,gg}(t)} \int_0^t\text{d}t_\text{B}\, \frac{\Gamma}{1-\mu}\rho_{ee}(t_\text{B})\left(1+\mu^2 \frac{\rho_{ee}(t_\text{B})}{\rho_{gg}(t_\text{B})}\right)\nonumber\\
=& A_0 \int_0^\infty\text{d}t\, \Gamma\exp(-2 \Gamma t) \exp\Big(A_2\, \exp(-\Gamma t)\Big) 
\int_0^t\text{d}t_\text{B}\,\Gamma\exp(-\Gamma t_\text{B}) \exp\Big(A_1\,\exp(-\Gamma t_\text{B})\Big)\nonumber \\
=& \frac{A_0}{A_1}\left\{\exp(A_2)\left[\frac{1-\exp(A_2)}{A_2^2} + \frac{\exp(A_2)}{A_2} \right] - \left[\frac{1 - \exp(A_1+A_2)}{(A_1+A_2)^2} +\frac{\exp(A_1+A_2)}{A_1+A_2}\right]\right\},
\label{Appeq:P_BA}
\end{align}
where we have defined
\begin{equation}
A_0 = \frac{\mu^5}{1-\mu}\frac{\pi_e^3}{\pi_g^2}\exp\left(\frac{-\mu}{1-\mu}\frac{\pi_e}{\pi_g}\right),\;\;\; A_1 = (\mu+\mu^2)\frac{\pi_e}{\pi_g},\;\;\; A_2 = \frac{\mu^3}{1-\mu}\frac{\pi_e}{\pi_g}.
\end{equation}

\end{document}